\documentclass{article}

\usepackage{arxiv}

\usepackage[utf8]{inputenc} 
\usepackage[T1]{fontenc}    
\usepackage{hyperref}       
\usepackage{url}            
\usepackage{booktabs}       
\usepackage{amsfonts}       
\usepackage{nicefrac}       
\usepackage{microtype}      
\usepackage{lipsum}
\usepackage{graphicx}
\graphicspath{ {./images/} }

\RequirePackage{amsthm,amsmath,amsfonts,amssymb}
\RequirePackage[numbers]{natbib}

\usepackage{algorithm,appendix}
\usepackage{algorithmic,multirow,booktabs,makecell}

\newtheorem{theorem}{Theorem}

\newtheorem{remark}{Remark}

\title{A novel framework for detecting multiple change points in functional data sequences}

\author{
 Zhiqing Fang \\
  School of Statistics and Management, \\
  Shanghai University of Finance and Economics
  \texttt{fang.zhiqing@163.sufe.edu.cn} \\
   \And
 Xin Liu \\
  School of Statistics and Management, \\
  Shanghai University of Finance and Economics
  \texttt{liu.xin@mail.shufe.edu.cn} \\
}

\begin{document}

\maketitle

\begin{abstract}
Detecting multiple change points in functional data sequences has been increasingly popular and critical in various scientific fields. In this article, we propose a novel two-stage framework for detecting multiple change points in functional data sequences, named as detection by Group Selection and Partial $F$-test (GS-PF). The detection problem is firstly transformed into a high-dimensional sparse estimation problem via functional basis expansion, and the penalized group selection is applied to estimate the number and locations of candidate change points in the first stage. To further circumvent the issue of overestimating the true number of change points in practice, a partial $F$-test is applied in the second stage to filter redundant change points so that the false discovery rate of the $F$-test for multiple change points is controlled. Additionally, in order to reduce complexity of the proposed GS-PF method, a link parameter is adopted to generate candidate sets of potential change points, which greatly reduces the number of detected change points and improves the efficiency. Asymptotic results are established and validated to guarantee detection consistency of the proposed GS-PF method, and its performance is evaluated through intensive simulations and real data analysis, compared with the state-of-the-art detecting methods. Our findings indicate that the proposed GS-PF method exhibits detection consistency in different scenarios, which endows our method with the capability for efficient and robust detection of multiple change points in functional data sequences.
\end{abstract}


\section{Introduction}
Statistical analysis of functional data sequences has become increasingly important in many scientific fields, including climatology \cite{shang2011nonparametric}, finance \cite{kokoszka2012functional}, geophysics \cite{hormann2012functional}, demography \cite{hyndman2008stochastic}, manufacturing \cite{Woodall2007CurrentRO}, and environmental modeling \cite{finazzi2019statistical, fortuna2020functional, qu2021robust}. Analogous to time series, a functional data sequence is observed and collected at each time point.
Just as univariate and multivariate time series, a functional data sequence can experience abrupt changes in its generating process, usually referred to as change points, which bring statistical challenges by violating stationarity assumptions when modeling in conventional times series analysis.

Within the functional data analysis (FDA) literature, change point detection has largely focused on the at-most-one-change (AMOC) problem, i.e., whether or not there will be a change point. Usually, a CUSUM-type test is employed for change point detection in a sequence of independent functional series \cite{berkes2009detecting}, and asymptotic theory for the corresponding estimator is established \cite{aue2009estimation}. Subsequently, the CUSUM-type test is further extended to accommodate functional data under a weakly dependent structure \cite{hormann2010weakly} and deals with epidemic changes in which the mean function shifts at some time point and then returns back \cite{aston2012detecting, zhang2011testing}. Besides, for the functional AMOC issue, employing a functional method without dimensionality reduction allows for the detection of mean changes with little information loss \cite{aue2018detecting}. More recently, statistical tools are developed for inferring changes of the patten in a spatiotemporal process, such as \cite{gromenko2017detection}. 

While these pioneering works may detect a single change point in a sequence of functional data with certain assumptions, they may fail in multiple change point detections, as the number and locations of multiple change points are difficult to estimate simultaneously, especially in high dimension and correlated scenarios. One way to deal with the problem is to transform multiple change points detection as a optimization problem where change points are usually regarded as an optimal set, assuming the number of change points is either known or unknown \cite{chiou2019identifying, pan2006application}, and then certain information criterion may be employed to estimate the total number of change points if necessary, such as the Schwarz criterion \cite{braun2000multiple, ciuperca2011general, yao1988estimating}. Alternatively, multiple change points can be identified using test-based methods. A popular approach is to create the CUSUM-type test statistic derived from the AMOC method, which is usually augmented with a recursive binary segmentation algorithm to partition the functional data sequences \cite{ aue2018detecting, berkes2009detecting}, and the detection consistency of using such a binary segmentation approach may be established \cite{rice2022consistency}. Similarly, a kernel-based functional seeded binary segmentation test is proposed, with its consistency validated for multiple change points estimation \cite{madrid2022change}, and a generic algorithm combines a suite of homogeneity tests conducted at multiple positions across diverse scales within data, capable of identifying significant change points under certain conditions \cite{pilliat2023optimal}. A third method to detect multiple change points is to utilize regression methods. From a regression perspective, multiple change points detection can be reformulated in a penalized selection context. For example, a penalized least-square criterion with a $\ell_1$-norm penalty is used for detecting change points with an almost optimal consistency rate asymptotically \cite{harchaoui2010multiple, huang2005detection}, such as the fused LASSO \cite{tibshirani2005sparsity}. An augmented fused lasso procedure is presented to split the projections of functional observations into multiple regions \cite{harris2022scalable} and then isolate each change point away from others so that the powerful univariate CUSUM statistic can be applied region-wisely. Additionally, the estimation problem of multiple-regime structural breaks is reformulated in an autoregressive model selection context \cite{chan2014group}, and
a dynamic sparse subspace learning approach is proposed for online structural change point detection of high-dimensional streaming data \cite{xu2023online}. Nonetheless, the regression-based method may only be feasible to roughly locate change points and tend to overestimate them, and filtering may still be necessary to detect the specific locations and number of change points. Another issue is that it may only detect change points on data with some specific structures, such as a notably large sample size between adjacent change points, but this may fail for complex data such as functional processes, and hence may be less accurate or robust to heavy tails in data distribution. Consequently, to the best of our knowledge, very limited research has focused on multiple change points detection for complex functional data sequences, and hence research gap still remains. 

Inspired by these statistical challenges, a novel framework is proposed to identify multiple change points in functional data sequences in this article, named as the Group Selection and Partial $F$-test (GS-PF) method. As a two-stage approach, the proposed method transforms the functional change point detection problem into a high-dimensional estimation context in the first stage, so that the number and locations of candidate change points will be estimated under certain assumptions, and then a partial $F$-test is subsequently used to filter redundant ones and control the false discovery rate in the second. To reduce the computational complexity of the GS-PF method, a link parameter is introduced to generate candidate sets for change points, significantly decreasing the number of candidate change points and thereby enhancing the efficiency of the partial $F$-test. Asymptotic results are established and justified to ensure the detection consistency. Numerical investigations on both simulated and real datasets have supported the superb performance of the proposed GS-PF method in different scenarios. Our method remedies three outstanding issues in the functional change point detection problem. Firstly, our method does not require prior knowledge of the number of change points, and even allows the number to increase with the sample size. Secondly, there is no prerequisite for a notably sample size between adjacent change points in our method, which enhances detection accuracy and consistency. Lastly, our method turns out to be robust to asymmetric and heavy tails in data distribution.

The rest of the article is organized as follows. In Section 2, the GS-PF framework of multiple change points detection for functional data sequences is introduced, and the estimation procedure is outlined. Theoretical justifications are presented in Section 3, and the performance of the proposed method is examined by simulation studies in Section 4 and real data analysis in Section 5. In Section 6, we conclude the article followed by a discussion.

\section{A novel GS-PF framework for change point detection in functional sequences}

In this section, we will introduce a novel framework of multiple change points detection in functional sequences, and then estimate the number and locations of change points using a two-stage approach, named as the Group Selection and Partial $F$-test (GS-PF) method. In the first stage, candidate change points are detected using group selection in a regression context, and a partial $F$-test on all candidate change points is implemented in the second stage under a false discovery rate (FDR) control so that the estimated change points are identified more accurately.

\subsection{A multiple change points model for functional sequences}\label{sec:change points model}
To start with, suppose that functions $\left\{f_t(x), x \in \left[0,1\right]; t \in \mathcal{T}=\{1, \ldots, T\} \right\}$ are generated from some distribution $F\left(\mu_t, \Sigma_t\right)$ with mean function $\mu_t$ and covariance function $\Sigma_t$ at time $t$. Specifically, we consider a multiple functional change points model
\begin{equation} \label{eq:cp model v0}
    f_t(x) = \mu_t(x) + \epsilon_t(x),
\end{equation}
where $f_t(x)$ is a continuous and square integrable function in $L^2([0,1])$ for a given $t$, satisfying $\int_0^1\left|f_t(x)\right|^2 d x<\infty$, and $\epsilon_t \sim \left(0, \Sigma_t \right)$. Without loss of generality, $f_t(x)$ is assumed to be observed on the same finite grid of points $0<x_1<\ldots<$ $x_d<1$ over $[0,1]$ for different $t$, and each function $f_t(x)$ is assumed piece-wise weakly stationary, i.e., $\mu_{t}=\mu_{t-1}$ for all $t \in \{1, \ldots, T\}$, except when $t$ is a change point. Let $\left\{\tau_m; m \in\{ 1, \ldots, M\} \right\}\subset \mathcal{T}$ denote locations of change points where $\mu_t$ changes, i.e., $\mu_{\tau_m} \neq \mu_{\tau_{m-1}}$. The goal of our method is to estimate change points' number $M$ and locations $\tau_1, \ldots, \tau_M$ from the observed sequences $\left\{f_t(x); x \in \{x_1, \ldots, x_d\}, t \in \{1, \ldots, T\right\}\}$.

However, to estimate change points using \eqref{eq:cp model v0} directly turns out to be difficult, since the functional sequences are only observed at discrete times, and reparameterization is accordingly employed. Specifically, let $\beta_1(x) = \mu_1(x)$, and $\beta_i(x) = \mu_i(x) - \mu_{i-1}(x)$ for $i \in \{2,3,\ldots,T\}$, and consequently \eqref{eq:cp model v0} is reorganized as
\begin{equation} \label{eq:cp model v1}
    f_t(x) = \sum_{i=1}^{t} \beta_i(x) + \epsilon_t(x).
\end{equation}
Further denoting $y_1(x) = f_1(x)$, $y_t(x) = f_t(x) - f_{t-1}(x)$, $\xi_1(x) = \epsilon_1(x)$, $\xi_t(x) = \epsilon_t(x) - \epsilon_{t-1}(x)$ for $t \in \{2,3,\ldots,T\}$, it is easily obtained that
\begin{equation} \label{eq:cp model v2}
    y_t(x) = \beta_t(x) + \xi_t(x).
\end{equation}
Consequently, detecting $\mu_{\tau_m} \neq \mu_{\tau_{m-1}}$ in \eqref{eq:cp model v0} is equivalent to detecting $\beta_t$ being a non-zero-valued function in \eqref{eq:cp model v2} when $t=\tau_m$. As $\beta_t(x)$ may not be directly observed in practice, it is necessary to employ its surrogate, achieved by expanding $\beta_t(x)$ over a certain set of basis functions $\{b_k(x)\}_{k=1}^K$ as
\begin{equation} \label{eq:basis expansion}
    \beta_t(x)
    = \sum_{k=1}^K a_{tk}b_k(x)
    := \boldsymbol{a}_t ^\top\boldsymbol{b}(x),
\end{equation}
where $\boldsymbol{b}(x) = (b_1(x), \ldots, b_K(x))^\top$, $\boldsymbol{a}_t = (a_{t1}, \ldots, a_{tK})^\top$ are time-varying unknown coefficient parameters, and $K$ is the number of basis functions that may be infinite theoretically. Hence, after plugging \eqref{eq:basis expansion} into \eqref{eq:cp model v2}, the identification of change points is also equivalent to detecting all non-zero $K$ dimensional parameter vectors $\boldsymbol{a}_t$ for $t \in \{2,\ldots, T\}$. In real implementation, a truncation may be employed on $K$ using certain information criterion.

\begin{remark}
    The basis in \eqref{eq:basis expansion} can be any kind of basis in $L^2$, such as the eigenfunction of the covariance operator of $f_t(x)$, the B-spline basis (such as cubic splines) and etc.. In this paper, the functional principal component analysis (FPCA) is considered.
\end{remark}

\subsection{The group selection (GS) stage}

Statistically, such an estimation problem can be addressed by employing a regression method, which minimizes the mean integrated squared error 
\begin{equation}\label{eq:isse}
    \tilde{L}(\boldsymbol{a}_t)
    =\frac{1}{T}\sum_{t=1}^{T}\int_0^1 \left(y_t(x) - \boldsymbol{a}_t^\top \boldsymbol{b}(x) \right)^2 d x
\end{equation}
over $\boldsymbol{a}_t$ for $t \in \{2,\ldots, T\}$. To control the complexity, two assumptions are usually set, i.e., the smoothness in the functionals $f_t(x)$ in \eqref{eq:cp model v0} and the number of change points. On one hand, as the smoothness of $f_t(x)$ is controlled by $\boldsymbol{b}^{\prime\prime}(x)$, the second-order derivative of $\boldsymbol{b}(x)$, a penalty $J_1(\boldsymbol{a}_t;\boldsymbol{b}^{\prime\prime}(x))$ over $\boldsymbol{a}_t$ is usually considered over \eqref{eq:isse}, which is the sum of (concave) penalty functions over $\boldsymbol{a}_t$ with a tuning parameter $\eta$, where a larger $\eta$ results in a smoother $\beta_t(x)$. 
On the other hand, the sparsity assumption when detecting change points is usually imposed that only a small set of change points exist in functional sequences, which is equivalent that only a small set of $\boldsymbol{a}_t$'s are non-zero. Consequently, a second penalty function $J_2(\boldsymbol{a}_t;\boldsymbol{b}(x))$ over $\boldsymbol{a}_t$ in \eqref{eq:isse} is introduced with a tuning parameter $\lambda$ controlling sparsity, where a larger $\lambda$ results in greater sparsity. Wrapping up $\tilde{L}(\boldsymbol{a}_t)$, $J_1(\cdot)$ and $J_2(\cdot)$, a penalized regression objective function of $\boldsymbol{a}_t$ is proposed 
\begin{align}\label{eq:pLa} 
    L(\boldsymbol{a}_t) 
    =& \tilde{L}(\boldsymbol{a}_t) + J_1(\boldsymbol{a}_t;\boldsymbol{b}^{\prime\prime}(x)) + J_2(\boldsymbol{a}_t;\boldsymbol{b}(x)) \nonumber \\
    =& \tilde{L}(\boldsymbol{a}_t) + \sum_{t=1}^{T} \rho_{1}\left(\eta \int_0^1 \left( \boldsymbol{a}_t \boldsymbol{b}^{\prime\prime}(x) \right)^2 d x\right)
    + \sum_{t=1}^{T} \rho_{2}\left(\lambda \int_0^1 \left( \boldsymbol{a}_t \boldsymbol{b}(x) \right)^2 d x\right),
\end{align}
and minimizing $L(\boldsymbol{a}_t)$ over $\boldsymbol{a}_t$ leads to the estimated non-zero $K$-dim $\boldsymbol{a}_t$. Note that if $\boldsymbol{a}_t$ is a non-zero vector, then $t$ is change point, but not vice versa.

Essentially, minimizing \eqref{eq:pLa} turns out to be a group selection problem.  Various penalty functions are available, such as the group LASSO and the group MCP. Specifically, while the group LASSO penalty $\rho\left(\left\|\boldsymbol{a}_t\right\|_2 ; \lambda\right)=\lambda \|\boldsymbol{a}_t\|_2$ may not guarantee consistency \cite{yuan2006model}, we consider the minimax concave penalty (MCP) $\rho\left(\|\boldsymbol{a}_t\|_2 ; \lambda, \gamma \right)=\lambda \int_0^{\|\boldsymbol{a}_t\|_2}(1-x /(\gamma \lambda))_{+} dx, \gamma>1$, where $\gamma$ is an nuisance parameter that controls the concavity of the penalty function $\rho(\cdot)$\cite{Zhang_2010}. The group MCP penalty induces sparsity and ensures consistent variable selection within groups. By approximating integrals using summation and integrating the two penalties into one, $L(\boldsymbol{a}_t)$ in \eqref{eq:pLa} is reformulated as
\begin{align}
    L(\boldsymbol{a}_t) 
    =& \left\|\mathbf{y}-\sum_{t=1}^{T}\mathbf{B}_t\boldsymbol{a}_t\right\|_2^2 + \sum_{t=1}^{T}\rho\left(\left\|\boldsymbol{a}_t\right\|_{R_t} ; \lambda, \eta, \gamma\right) \label{eq:optimiation v3}.
\end{align}
where
$\mathbf{y}=(y_1(x_1),y_1(x_2),\ldots,y_1(x_d),y_2(x_1),y_2(x_2),\ldots,y_2(x_d),\ldots,y_T(x_1 ),y_T(x_2),\ldots, \\ y_T(x_d))^\top$ is a $Td \times 1$ vector,
$\mathbf{B}_t = 
\begin{bmatrix}
    \mathbf{O},\ldots ,\mathbf{O},\boldsymbol{b}(\boldsymbol{x}),\mathbf{O},\ldots,\mathbf{O}
\end{bmatrix}^\top$ is a $Td \times K$ matrix whose elements are $K \times d$ zero matrices except for the $t$- element as $\boldsymbol{b}(\boldsymbol{x})$, $\left|\left|\cdot\right|\right|$ is $l_2$-norm, $\rho(\cdot)$ is the penalty function, $\left|\left|\boldsymbol{a}_t\right|\right|_{\mathbf{R}_t} = \sqrt{\boldsymbol{a}_t^{\top}\mathbf{R}_t\boldsymbol{a}_t}$, $\mathbf{R}_t = \eta \boldsymbol{b}(\boldsymbol{x})^{\prime \prime}\boldsymbol{b}(\boldsymbol{x})^{\prime \prime \top} + \lambda \boldsymbol{b}(\boldsymbol{x})\boldsymbol{b}(\boldsymbol{x})^{\top}$ is a $K \times K$ positive definite matrix,
$\boldsymbol{b}^{\prime \prime}(\boldsymbol{x})=
\begin{bmatrix}
    \boldsymbol{b}_1^{\prime \prime}(\boldsymbol{x}), \boldsymbol{b}_2^{\prime \prime}(\boldsymbol{x}), \ldots, 
    \boldsymbol{b}_K^{\prime \prime}(\boldsymbol{x})
\end{bmatrix}^\top$
is a $K \times d$ matrix with $\boldsymbol{b}_k^{\prime \prime}(\boldsymbol{x})=(b_1^{\prime \prime}(x_1), b_1^{\prime \prime}(x_2), \ldots, b_1^{\prime \prime}(x_d))^\top$,
$\boldsymbol{b}(\boldsymbol{x})=
\begin{bmatrix}
    \boldsymbol{b}_1(\boldsymbol{x}), \boldsymbol{b}_2(\boldsymbol{x}), 
    \ldots, 
    \boldsymbol{b}_K(\boldsymbol{x})
\end{bmatrix}^\top$
is a $K \times d$ matrix with $\boldsymbol{b}_k=(b_1(x_1), b_1(x_2), \ldots, b_1(x_d))^\top$, and $\lambda$ and $\eta$ are tuning parameters. By Cholesky decomposition, there exists a $K \times K$ upper triangular matrix $\mathbf{L}$ such that $\mathbf{R}_t=\mathbf{L}^T\mathbf{L}$. Let $\boldsymbol{\alpha}_t=\mathbf{L}\boldsymbol{a}_t$, then $\boldsymbol{a}_t=\mathbf{L}^{-1}\boldsymbol{\alpha}_t$ and $\|\boldsymbol{a}_t\|_{R_t} = \|\boldsymbol{\alpha}_t\|_2$. Consequently, we can transform \eqref{eq:optimiation v3} into 
\begin{align} \label{eq:optimiation v4}
    L(\boldsymbol{\alpha}) = \left\|\mathbf{y}-\sum_{t=1}^{T}\mathbf{B^{\ast}_t}\boldsymbol{\alpha}_t\right\|_2^2 + \sum_{t=1}^{T}\rho\left(\left\|\boldsymbol{\alpha}_t\right\|_2 ; \lambda, \eta, \gamma \right).
\end{align}
where $\mathbf{B^{\ast}_t}=\mathbf{B}_t\mathbf{L}$. By estimating $\boldsymbol{\alpha}_t$ in \eqref{eq:optimiation v4} and further $\boldsymbol{a}_t$ in \eqref{eq:optimiation v3}, we can judge whether $\beta_t(x)$ in \eqref{eq:basis expansion} is a non-zero valued function and then get the candidate change points, i.e., elements of the set $A(\hat{\boldsymbol{\alpha}}_t)=\left\{t: \hat{\boldsymbol{\alpha}}_t \neq \boldsymbol{0} \right\}$ for a given $\lambda$ and $\eta$. If $t$ corresponds to a zero-valued function $\hat{\beta}_t(x)$, then $t$ is a change point, which means $\hat{\mu}_{t} \neq \hat{\mu}_{t-1}$ in \eqref{eq:cp model v0}. The estimator $\hat{\mu_{t}}$ will necessarily be a piece-wise constant vector, due to the penalty term 
$\rho\left(\left\|\boldsymbol{\alpha}_t\right\|_2 ; \lambda, \eta, \gamma\right)$ that induces sparsity on the differences of $\mu_{t}$ \cite{Zhang_2010}.

\begin{remark}
    Note that our model does not impose any restriction between the sampling frequency $d$ and the number of functional curves $T$. Indeed, our approach can handle both the dense case where $d$ exceeds $T$ with the same magnitude of group selection technology and the sparse case where $d$ can be upper bounded by a constant. Besides the fixed sampling scheme studied here, another possibly studied scenario is the random design, where the sampling locations are assumed to be mutative to functional curves across time. Although we focus on the fixed design here, our model can be applied to the random case through a reasonable transformation, by firstly fitting the discrete observation points into a continuous curve and then taking the function value at fixed points.
\end{remark}

\subsection{The partial $F$-test (PF) stage}
To further refine the candidate set $A(\hat{\boldsymbol{\alpha}}_t)=\left\{t: \hat{\boldsymbol{\alpha}}_t \neq \boldsymbol{0} \right\}$ from the GS stage and accurately locate change points, we consider statistical tests to control the errors and we employ the partial $F$-test to ascertain if a candidate change point can be finally determined as a change point in particular. Usually, the partial $F$-test assesses whether additional variables collectively contribute enough explanatory power to justify their inclusion in the equation and it essentially compares whether the full model significantly outperforms a reduced version \cite{kleinbaum1988applied}. Specifically in our model setting, the partial $F$-test is implemented as follows. 

Consider a standard multiple linear regression model given by
\begin{equation} \label{eq:full model}
    \mathbf{y}= \mathbf{B^{\ast}}\boldsymbol{\alpha}+\mathbf{\xi},
\end{equation}
where 
\begin{itemize}
    \item $\mathbf{y}=(y_1(x_1),y_1(x_2),\ldots,y_1(x_d),\ldots,y_T(x_1),y_T(x_2),\ldots,y_T(x_d))^\top$ is a $Td \times 1$ vector of observations, 
    \item $\mathbf{B^{\ast}} = 
    \begin{bmatrix}
        \boldsymbol{1},
        \mathbf{B^{\ast}_1},
        \mathbf{B^{\ast}_2},
        \ldots,
        \mathbf{B^{\ast}_T}
    \end{bmatrix}$ is an $Td \times (TK+1)$ full column-rank design matrix with the first column given by $(1, \ldots, 1)^{\mathrm{T}}$ with length $Td$, 
    \item $\boldsymbol{\alpha}=(\alpha_0,\alpha_1(x_1),\alpha_1(x_2),\ldots,\alpha_1(x_k),\ldots,\alpha_T(x_1),\alpha_T(x_2),\ldots,\alpha_T(x_k))^\top
    $ is a $(Tk+1) \times 1$ vector of unknown coefficients with a constant coefficient $\alpha_0$,
    \item $
    \mathbf{\xi}=(\xi_1(x_1),\xi_1(x_2),\ldots,\xi_1(x_d),\ldots,\xi_T(x_1),\xi_T(x_2),\ldots,\xi_T(x_d))^\top
    $ is a $Td \times 1$ vector of random errors.
\end{itemize}

\noindent To eliminate the heteroskedasticity by the reparameterization in \eqref{eq:cp model v2}, \eqref{eq:full model} is further transformed into \eqref{eq:transformed full model}
\begin{equation} \label{eq:transformed full model}
    \mathbf{\tilde{y}}= \mathbf{\tilde{B}^{\ast}}\boldsymbol{\alpha} + \mathbf{\tilde{\xi}},
\end{equation}
where $\mathbf{\tilde{y}}=\Sigma_{\xi}^{-\frac{1}{2}} \mathbf{y}$, $\mathbf{\tilde{B}^{\ast}}=\Sigma_{\xi}^{-\frac{1}{2}}\mathbf{B^{\ast}}$, $\mathbf{\tilde{\xi}}=\Sigma_{\xi}^{-\frac{1}{2}}\mathbf{\xi}$, and $\Sigma_{\xi}$ is the covariance matrix of $\mathbf{\xi}$. The key problem in \eqref{eq:transformed full model} is to assess whether some of the coefficients are zero and so the corresponding covariates in $\mathbf{\tilde{B}^{\ast}}$ have no effect on the response variable $\mathbf{y}$. To illustrate, let $\boldsymbol{\alpha}=\left(\boldsymbol{\alpha}^{(1)\top}, \boldsymbol{\alpha}^{(2)\top}\right)^\top$, where $\boldsymbol{\alpha}^{(1)}=\left(\alpha_0, \ldots, \alpha_{p-k}\right)^\top$ and $\boldsymbol{\alpha}^{(2)}=\left(\alpha_{p-k+1}, \ldots, \alpha_p\right)^\top$ with $1 \leqslant k \leqslant p$ and $p=Tk+1$. If $\boldsymbol{\alpha}^{(2)}$ is truly zero, then the corresponding covariates have no effect on the response variable $\mathbf{\tilde{y}}$, and  \eqref{eq:transformed full model} reduces to
\begin{equation} \label{eq:reduced model}
    \mathbf{\tilde{y}}=\mathbf{\tilde{B}^{\ast}}_1 \boldsymbol{\alpha}^{(1)}+\mathbf{\tilde{\xi}}_1
\end{equation}
where $\mathbf{\tilde{B}^{\ast}}_1$ is formed by the first $p-k+1$ columns of the matrix $\mathbf{\tilde{B}^{\ast}}$.
A commonly used statistical approach to assessing whether $\boldsymbol{\alpha}^{\left(2\right)}$ a zero vector is to test the hypotheses
\begin{equation} \label{test:partial $F$-test}
    \mathrm{H}_0: \boldsymbol{\alpha}^{(2)}=\mathbf{0} \quad v.s. \quad \mathrm{H}_1: \boldsymbol{\alpha}^{(2)} \neq \mathbf{0}
\end{equation}
by using the partial $F$-test, which rejects $\mathrm{H}_0$ if and only if
$$
F = \frac{(\text{SSR of } \eqref{eq:full model}-\text{SSR of } \eqref{eq:reduced model}) / k}{\text {MSE of \eqref{eq:full model}}} > F_\alpha(k, n-p-1),
$$
where SSR denotes the sum of squares of regression, MSE denotes the mean squared error, $F_\alpha(k, n-p-1)$ is the upper $100\alpha\%$-quantile of an $F$-distribution with its degrees of freedom as $k$ and $n-p-1$, respectively.

Specifically in our setting, such a partial $F$-test is employed to examine a candidate change point $m \in A(\hat{\boldsymbol{\alpha}}_t):=\left\{t: \hat{\boldsymbol{\alpha}}_t \neq \boldsymbol{0} \right\}$ from the first stage, and it is equivalent to test whether $\boldsymbol{c}_m \in \{\hat{\boldsymbol{\alpha}_t}: \hat{\boldsymbol{\alpha}}_t \neq \boldsymbol{0} \}$ is a zero vector. In this case, $\boldsymbol{\alpha}$ in the full model \eqref{eq:transformed full model} represents a long vector formed by concatenating the non-zero vectors $\hat{\boldsymbol{\alpha}}_t$ corresponding to all the change points $t \in A(\hat{\boldsymbol{\alpha}}_t)$ and the zero vectors $\hat{\boldsymbol{\alpha}}_t$ corresponding to all the non-change points $t \in \bar{A}(\hat{\boldsymbol{\alpha}}_t):=\left\{t: \hat{\boldsymbol{\alpha}}_t = \boldsymbol{0} \right\}$, while in the reduced model \eqref{eq:reduced model}, $\boldsymbol{\alpha}^{(1)}$ represents a long vector formed by concatenating the non-zero vectors $\hat{\boldsymbol{\alpha}}_t$ corresponding to all the change points $t \in \left\{t: \hat{\boldsymbol{\alpha}}_t \neq \boldsymbol{0} \text{ and } \hat{\boldsymbol{\alpha}}_t \neq \boldsymbol{c}_m \right\}$ and the zero vectors $\hat{\boldsymbol{\alpha}}_t$ corresponding to all the non-change points $t \in \bar{A}(\hat{\boldsymbol{\alpha}}_t):=\left\{t: \hat{\boldsymbol{\alpha}}_t = \boldsymbol{0} \right\}$, and $\boldsymbol{\alpha}^{(2)}$ in \eqref{test:partial $F$-test} is $\boldsymbol{c}_m$. The decision to keep or remove $m$ from $A(\hat{\boldsymbol{\alpha}}_t)$ is based on the outcome of the partial $F$-test: if the null hypothesis $\mathrm{H}_0$ in \eqref{test:partial $F$-test} is rejected, the candidate change point $m$ is retained; otherwise, it is removed. 

\begin{remark}
    The specific form of the covariance matrix $\Sigma_{\xi}$ depends on $\Sigma_t$ in Section \ref{sec:change points model}. For a typical $\Sigma_t$, it can be proved that $\Sigma_{\xi}$ is invertible. In cases where an invertible matrix $\Sigma_{\xi}$ cannot be constructed, the transformation remains valid through the use of the generalized inverse. During the actual execution of the partial $F$-test, the sample covariance matrix of $\Sigma_{\xi}$ is utilized, serving as a consistent estimator.
\end{remark}

\subsection{The implementation of the GS-PF method}
In the proposed GS-PF method, four hyper-parameters are involved, i.e., the MCP penalty term $\lambda, \eta$ and $\gamma$ and the false discovery rate (FDR) $\alpha$. We set $\gamma=3$ as suggested in the literature \cite{liu2019subgroup,ma2017concave,yang2019high,zhang2016gamma} and use data to optimize $\lambda$ and $\eta$ parameter through grid search based on BIC minimization. 
The FDR $\alpha$, the expected ratio of the number of false change points over the number of detected change points, is applied to control the number of false detected change points at a specified level $\alpha$. 
Typically, The FDR remains fixed in a standard hypothesis testing, although this parameter may be optimized as well as $\gamma$.

Further, as each candidate change point is tested independently, there will be $\hat{M}$ tests in total in the PF stage after $\hat{M}$ candidate change points are detected in the GS stage. To further control the FDR at a predetermined significance level $\alpha$, the Benjamini-Hochberg (BH) procedure \cite{benjamini1995controlling} is employed to adjust the $\hat{M}$ $p$-values, in line with Theorem \ref{th:FDR using BH}. Accordingly, all candidate change points with an adjusted $p$-value below the prespecified $\alpha$ level will be retained. The whole procedure of the GS-PF method is wrapped up in Algorithm \ref{algor:GS-PF}. 

\begin{algorithm}
\caption{The GS-PF Algorithm}
\label{algor:GS-PF}
\begin{algorithmic}[1]
\STATE \textbf{Input: }$\{f_t\}_{t=1}^T, \alpha$
\STATE Use group selection to estimate $\boldsymbol{\alpha}_t$ and select $A(\hat{\boldsymbol{\alpha}}_t)=\left\{t: \hat{\boldsymbol{\alpha}}_t \neq \boldsymbol{0} \right\}$ to obtain candidate change points by tuning the parameters $\lambda$, $\eta$ and $\gamma$
\STATE Do partial $F$-test on candidate change points and calculate adjusted $p$-values using BH procedure
\IF {adjusted $p$-value $\leq \alpha$}
    \STATE elevate the candidate change point to a change point
\ELSE
    \STATE abandon the candidate change point
\ENDIF
\STATE \textbf{Output: } change points
\end{algorithmic}
\end{algorithm}

\begin{remark}
    It seems we could simply identify change points with the GS-PF method and the group selection is consistent theoretically. In practice, however, the set $A(\hat{\boldsymbol{\alpha}})$ tends to overestimate the number of change points and sometimes estimates a single change point with a sequence of closely grouped estimates around the true value. Performing a partial $F$-test for each candidate change point would escalate computational demands due to potentially substantial overestimation, thus increasing the computational burden. Therefore, we need a technique to reduce the time complexity of the calculation. To carry out the refinement, we generate the candidate sets from the candidate change points identified through group selection. This is achieved by adopting the link parameter $\varkappa$, which represents the maximum distance between any two neighboring candidate change points that are presumed to correspond to a common actual change point \cite{harris2022scalable}. We amalgamate the elements of $A(\hat{\boldsymbol{\alpha}})$ into sets, that are at least $\varkappa+1$ time steps apart, by linking together candidate change points that are within $\varkappa$ time steps of each other. The link parameter $\varkappa$ are optimized simultaneously with the parameter $\lambda$ and $\eta$ through grid search based on BIC minimization. If the candidate set contains only one element, we keep it as a change point representative. Otherwise, we employ functional CUSUM statistics within the candidate set and elect a change point representative under an AMOC model. Finally, we merely need to apply the partial $F$-test to change point representatives, all change point representatives with an adjusted p-value below the pre-specified $\alpha$ level using the BH method will be regarded as the estimated change-points. Accordingly, a refined version of the GS-PF algorithm is proposed in Algorithm \ref{algor:GS-PF(refined version)}.
\end{remark}

\begin{algorithm}
\caption{GS-PF (refined version)}
\label{algor:GS-PF(refined version)}
\begin{algorithmic}[1]
\STATE \textbf{Input: }$\{f_t\}_{t=1}^T, \alpha$
\STATE Group selection to estimate $\boldsymbol{\alpha}_t$ and select $A(\hat{\boldsymbol{\alpha}}_t)=\left\{t: \hat{\boldsymbol{\alpha}}_t \neq \boldsymbol{0} \right\}$ to obtain candidate change points by tuning the parameters $\lambda$, $\eta$ and $\gamma$
\STATE Generate a series of candidate sets by tuning the link parameter $\varkappa$
\IF {a candidate set contains only one element}
    \STATE keep the element as a change point representative
\ELSE
    \STATE elect the candidate change point with the maximal functional CUSUM statistics as a change point representative
\ENDIF
\STATE Do partial $F$-test on change point representatives and calculate adjusted $p$-values using BH procedure
\IF {adjusted $p$-value $\leq \alpha$}
    \STATE elevate the change point representative to a change point
\ELSE
    \STATE abandon the change point representative
\ENDIF
\STATE \textbf{Output: } change points
\end{algorithmic}
\end{algorithm}

\section{Theoretical results}
\subsection{Detection consistency}
\noindent Before demonstrating the detection consistency of the proposed method, several assumptions are necessary and introduced below.

\begin{itemize}
    \item[(A1)] The noise vector $\boldsymbol{\epsilon} =(\epsilon_1(x_1),\epsilon_1(x_2),\ldots ,\epsilon_1(x_d),\epsilon_2(x_1),\epsilon_2(x_2),\ldots,\epsilon_2(x_d),\ldots,\epsilon_T(x_1 ), \\ \epsilon_T(x_2),\ldots,\epsilon_T(x_d))^\top$ in \eqref{eq:cp model v0} has sub-Gaussian tails such that 
    \begin{equation*}\label{eq:subG tail}
        \operatorname{Pr}\left(\left|\mathbf{v}^{\top} \boldsymbol{\epsilon}\right|>\|\mathbf{v}\|_2 s\right) \leq 2 \exp \left(-C s^2\right)
    \end{equation*}
    for any vector $\mathbf{v} \in \mathbb{R}^{Td}$  and $s>0$, where $0<C<\infty$.
    
    \item[(A2)] Let $\mathbf{X}=(\mathbf{B^{\ast}_1},\mathbf{B^{\ast}_2},\ldots,\mathbf{B^{\ast}_T})$, where $\mathbf{B^{\ast}_1},\mathbf{B^{\ast}_2},\ldots,\mathbf{B^{\ast}_T}$ are in \eqref{eq:optimiation v4} and $n=Td$, $\mathbf{\Sigma}=\mathbf{X}^\top\mathbf{X}/n$, and $c_{min}$ is the smallest eigenvalue of $\mathbf{\Sigma}$. Assume $\gamma \geq c_{min}^{-1}$.
    
    \item[(A3)] Denote the true value of $\boldsymbol{\alpha}$ as $\boldsymbol{\alpha}^0=(\boldsymbol{\alpha}^{0\top}_1,\boldsymbol{\alpha}^{0\top}_2,\ldots,\boldsymbol{\alpha}^{0\top}_T)^\top$, $A(\tau)=\{t: \|\boldsymbol{\alpha}^0_t\|_2 \neq 0, 1 \leq t \leq T\}$, and $$
    \boldsymbol{\alpha}^0_\ast=
    \begin{cases}
        \min\limits_{t \in A(\tau)}\left\{\frac{\|\boldsymbol{\alpha}^0_j\|_2}{\sqrt{d}}\right\} & \text { if } |A(\tau)|>0 \\ \infty & \text { if } |A(\tau)|=0
    \end{cases}.
    $$ Assume $\boldsymbol{\alpha}^0_\ast = \gamma\lambda\eta + o(n^{-q})$ with $\gamma$, $\lambda$ and $\eta$ in \eqref{eq:optimiation v4}, and $\lambda\eta=o(n^{-q})$ for $ q\in[0,1].$
\end{itemize}

\begin{remark}
    (A1) is a common condition in functional data analysis controlling the tail behavior. (A2) ensures the convexity of the group MCP, which implies that $\{\boldsymbol{\alpha}_t\}_{t=1}^T$ is uniquely characterized by the Karush–Kuhn–Tucker conditions. (A3) requires values of functional sequences before and after a change point cannot be too small.
\end{remark}

\noindent Under these assumptions, the detecting consistency of the change points by group selection method is given in Theorem \ref{th:consistency}.

\begin{theorem}\label{th:consistency}
    Denote $M=|A(\tau)|$ and $\hat{M}=|A(\hat{\boldsymbol{\alpha}}_t)|$, where $|A(\tau)|$ is defined in Assumption (A3) and $\hat{\boldsymbol{\alpha}}_t$ is the estimate of $\boldsymbol{\alpha}_t$ by minimizing $L(\boldsymbol{\alpha})$ over $\boldsymbol{\alpha}_t$ in \eqref{eq:optimiation v4}. Then under Assumptions (A1-A3), 
    \begin{equation}\label{eq:number consistency}
        \Pr(\hat{M}=M) \rightarrow 1, 
    \end{equation}
    and further
    \begin{equation}\label{eq:location consistency}
        \Pr(d_{H}(A(\hat{\boldsymbol{\alpha}}_t), A(\tau))=0) \rightarrow 1. 
    \end{equation}
    where $d_{H}(\cdot, \cdot)$ is the Hausdorff distance that measures how far away two subsets in a metric space are from each other \cite{rockafellar2009variational}.
\end{theorem}

\noindent Detailed proof of Theorem \ref{th:consistency} is provided in the appendix. Theorem \ref{th:consistency} implies that the proposed method can theoretically estimate the number $M$ and locations $\tau_1, \ldots, \tau_M$ of change points with probability approaching to one under some general assumptions.

\subsection{The FDR control using partial $F$-test}
\begin{theorem}\label{th:FDR using BH}
    For multiple hypothesis testing problem
    $$
    H_{0, m}: \boldsymbol{c}_m = \boldsymbol{0} \qquad v.s. \qquad H_{1, m}: \boldsymbol{c}_m \neq \boldsymbol{0}, \qquad \boldsymbol{c}_m \in \{\hat{\boldsymbol{\alpha}_t}: \hat{\boldsymbol{\alpha}}_t \neq \boldsymbol{0} \},
    $$
    then
    \begin{equation}\label{eq:location consistency}
        \Pr(FDR \leq \alpha) \rightarrow 1. 
    \end{equation}
\end{theorem}

\noindent Detailed proof of Theorem \ref{th:FDR using BH} is provided in the appendix.

\section{Simulation studies}
\subsection{Simulation Setup}
In this section, the proposed GS-PF method is carefully examined on functional sequences with different structures in various scenarios of change point detection, including no change point, one change point, and multiple change points, respectively. Specifically, five functional sequences are generated using the model $f_t^i(x) = \mu_t^i(x) + \epsilon_t^i(x)$, where $\mu_t^i(x) \in L^2([0,1])$, $i=1,2,3,4,5$. These sequences include constant, symmetric, asymmetric, piece-wisely stationary with a dyadic structure as well as a benchmark functional sequences, and the structures of $\mu_t^i(x)$ or $f_t^i(x)$ are listed in Table \ref{tab:function settings}. 

\begin{table}[h!]
    \centering
    \small
    \caption{Functional settings in five synthetic datasets}\label{tab:function settings}
    \begin{tabular}{m{1.8cm}|m{10.5cm}|m{2.8cm}}
        \hline
        \multicolumn{1}{c|}{Type} & \multicolumn{1}{c|}{Functions or mean functions} & \multicolumn{1}{c}{Error functions} \\
        \hline
        Constant & 
        $\begin{aligned}
            \mu_t^1(x) &= 0, 
            \mu_t^2(x) = 5, 
            \mu_t^3(x) = 7, 
            \mu_t^4(x) = 11,
            \mu_t^5(x) = 8
        \end{aligned}$ 
        & 
        $\epsilon_t^i(x)\sim$ a zero-mean GP with Matérn covariance function in \eqref{eq:matern cov} \\
        \hline
        Symmetric & 
        $\begin{aligned}
            \mu_t^1(x) &= 5 x^2 - \exp(1-20 x), \\
            \mu_t^2(x) &= -1 - 100(x-0.1)(x-0.3)(x-0.5)(x-0.9), \\
            \mu_t^3(x) &= \mu_t^2(x) - 2|\sin (1+10 \pi x)|, \\
            \mu_t^4(x) &= 1 + 3 x^2 - 5 x^3 - \sin (1+10 \pi x), \\
            \mu_t^5(x) &= 3 x^2 - 5 x^3
        \end{aligned}$ 
        &
$\epsilon_t^i(x)\sim$ a zero-mean GP with Matérn covariance function in \eqref{eq:matern cov} \\
        \hline
        Asymmetric &
        $\begin{aligned}
            f_t^i(x) &= \log \left(1+e^{g_t^i(x)}\right), 
        \end{aligned}$ where $g_t^i(x)$ equal to $f_t^i(x)$ in symmetric case, $i = 1, 2, 3, 4, 5$.
        & 
        $\epsilon_t^i$ is included in $f_t^i(x)$ \\
        \hline
        Piece-wise stationary process with dyadic structure &
        $\begin{aligned}
            f_t^1 &= 0.9 f_{t-1}^1 + \epsilon_t^1, \\ 
            f_t^2 &= 1.32 f_{t-1}^2 - 0.81 f_{t-2}^2 + 2 + \epsilon_t^2, \\
            f_t^3 &= -0.5 f_{t-1}^3 + 0.1 f_{t-2}^3 + 1 + \epsilon_t^3, \\
            f_t^4 &= 0.9 f_{t-1}^4 + \epsilon_t^4, \\ 
            f_t^5 &= 1.32 f_{t-1}^5 - 0.81 f_{t-2}^5 + 2 + \epsilon_t^5, 
        \end{aligned}$ 
        & $\epsilon_t^i \overset{\text{iid}}{\sim} \mathcal{N}(0,1)$ \\
        \hline
        Benchmark &
        $\begin{aligned}
            \mu_t^1(x) = 0, 
            \mu_t^2(x) = 3t, 
            \mu_t^3(x) = 6-2t^2, 
            \mu_t^4(x) = e^{at}, 
            \mu_t^5(x) = 7t^3
        \end{aligned}$ 
        & $\epsilon_t^i(x) \overset{\text{iid}}{\sim} \mathcal{N}(0,1)$ \\
        \hline
    \end{tabular}
\end{table}

Notably for datasets with error functions generated by Gaussian Processes (GP), the Matérn covariance function \cite{stein1999interpolation} is employed and defined as
\begin{equation}\label{eq:matern cov}
    C\left(x, x^{\prime}\right)=\frac{\sigma^2 \sqrt{\pi} r^{2 \nu}}{2^{\nu-1} \Gamma(\nu+1 / 2)}\left(\frac{\left\|x-x^{\prime}\right\|}{r}\right)^\nu K_\nu\left(\frac{\left\|x-x^{\prime}\right\|}{r}\right)
\end{equation}
with variance parameter $\sigma^2$, range parameter $r$, smoothness parameter $\nu$, and $K_\nu(\cdot)$ which represents a modified Bessel function of the second kind of order $\nu$, where $\sigma=0.1$ and $r=0.1$ are assigned for simplicity and $\nu=1$ is selected to ensure mean square differentiability of the sample paths so that the functional principal components are meaningful. Light-tailed and heavy-tailed functional data are simulated using GP and t-process (TP) models, respectively. Compared to the GP, the TP model requires an additional degrees of freedom parameter $df$, which is set to 3 in all simulation studies. To generate asymmetric functional sequences $\left\{f_t^i(x), t = 1, \ldots, n; i = 1, 2, 3, 4, 5 \right\}$, following the data generation mechanism \cite{harris2022scalable}, the transformation $f_t^i(x)=\log\left(1+e^{g_t^i(x)}\right)$ is applied to functional sequences $g_t^i(x)$, which are equivalent to $f_t^i(x)$ in symmetric case. Further, Among the various types of locally stationary models, the so-called structural-break or change point models has received particular attention. One particularly useful locally stationary model for describing structural-break or change point behavior is the so-called (m+1)-regime structural break autoregressive (SBAR) model. Following the data generation mechanism \cite{chan2014group}, the piece-wise stationary process with dyadic structure is generated. Note that this may be a tough scenario, as the signal to noise may not be sufficiently strong to detect change points with piecewise stationarity. Finally, in order to evaluate the proposed GS-PF method against state-of-the-art functional change point detection methods, the benchmark dataset is replicated from \cite{chi2023group} to ensure a fair comparison under identical experimental conditions.

Without loss of generality, the mean function of $f_t$ in the piece-wise stationary process with dyadic structure is also noted by $\mu_t$, then differences yield different scales of change between the mean functions $\mu_t$ of functions $f_t$, as is measured by $L_2$ norm. The changes are large from $\mu_t^1$ to $\mu_t^2$ and $\mu_t^5$ to $\mu_t^1$, moderate from $\mu_t^3$ to $\mu_t^4$, and relatively small from $\mu_t^2$ to $\mu_t^3$ and $\mu_t^4$ to $\mu_t^5$. Additionally, these $\mu_t$ functions represent changes in both magnitude and the shape of the functional process. Further, in order to generate $M$ randomly spaced change points of i.i.d. sequences, $M+1$ segment lengths (i.e., segment sample sizes) are randomly sampled from U$(100, 200)$, denoted as $n_1, \ldots, n_{M+1}$. Each segment is concatenated together so that the boundaries between segments represent change points in the functional data sequences. 

Further to generate different scenarios of change points, we consider situations of no change point, one change point and multiple change points, respectively. For the no change point setting, all functional sequences are generated only with $\mu_t^1$ or $f_t^1$ in Table \ref{tab:function settings}. For the one change point setting, functional observations are generated from $f_t^1$ first and then from $f_t^2$, indicating that a true change takes place from $\mu_t^1$ to $\mu_t^2$. For the multiple change point setting, we specify the number of true change points as $M=5$, with each change point occurring sequentially in the sequence comprising $f_t^i, i=1,2,3,4,5$ across the respective five datasets.

To evaluate the performance, the proposed GS-PF method is examined and compared with a fair competing method, namely the GLASSO method \cite{chi2023group} in all simulated datasets using two popular statistical measurements.
One is the annotation error that quantifies the difference in the number of detected change points versus the number of true change points \cite{truong2020selective}, and the other is the Hausdorff distance that measures the location deviation between the set of estimated change points and the set of true change points \cite{rockafellar2009variational}. Specifically in our case, let $A=\left\{a_1, \ldots, a_n\right\}$ be the set of the estimated change points, and $B=\left\{b_1, \ldots, b_m\right\}$ the set of the true change points, the annotation error between $A$ and $B$ is defined as
$d_A(A, B)=|n-m|$ and the Hausdorff error as
$d_H(A, B)=\max _{b \in B} \min _{a \in A}|b-a|$. A low annotation error indicates that the algorithm consistently identifies the correct number of change points, while a low Hausdorff error indicates strong similarity in location between estimated and true change points. Different detectors are compared based on their ability to minimize annotation errors and Hausdorff errors. In all simulations, detection is considered successful only when both the annotation error and the Hausdorff error are zero; otherwise, the detection fails. Particularly when one of $A$ and $B$ is an empty set $\emptyset$, $d_H(A, \emptyset)=d_H(\emptyset, B)=1$ \cite{Boysen_2009}.
Each experiment is repeated 100 times, and the average failure and success rates are summarized and reported respectively.

\subsection{Main findings in different scenarios}
To start with, the detection results for the no-change-point scenarios are summarized in Table \ref{tab:success rates M=0}. For the GS-PF method, the detection success rates increase as $\alpha$ decreases, eventually approaching to one, which indicates that the GS-PF method effectively boosts the detection strength by adjusting $\alpha$. When $\alpha=0.00001$, the detection success rates for all datasets rise above $95\%$. In contrast, the GLASSO method exhibits constant detection success rates, irrespective of $\alpha$ adjustments. Notably, the success rate for the benchmark dataset remains at $0$, suggesting that the GLASSO method cannot guarantee the success rate. Furthermore, the GS-PF method maintains excellent and consistent detection performance across all five types of mean functions, demonstrating its robustness.

\begin{table}[t]
\tiny
\centering
\caption{Success rates of change point detection with $M=0$ for the two methods under different datasets and $\alpha$ values.}
\label{tab:success rates M=0}
\begin{tabular}{ccccccccccc}
\hline
\multirow{2}{*}{$\alpha$} & \multicolumn{2}{c}{Constant} & \multicolumn{2}{c}{Symmetric} & \multicolumn{2}{c}{Asymmetric} & \multicolumn{2}{c}{Piece-wisely stationary} & \multicolumn{2}{c}{Benchmark} \\ \cline{2-11} 
& GS-PF & GLASSO & GS-PF & GLASSO & GS-PF & GLASSO & GS-PF & GLASSO & GS-PF & GLASSO \\ \hline
0.05                   & 0.57 & 1 & 1 & 1 & 1 & 1 & 0.82 & 0.8 & 0.56 & 0 \\
0.01                   & 0.69 & 1 & 1 & 1 & 1 & 1 & 0.88 & 0.8 & 0.62 & 0 \\
0.005                  & 0.75 & 1 & 1 & 1 & 1 & 1 & 0.89 & 0.8 & 0.64 & 0 \\
0.001                  & 0.86 & 1 & 1 & 1 & 1 & 1 & 0.97 & 0.8 & 0.70 & 0 \\
0.0005                 & 0.89 & 1 & 1 & 1 & 1 & 1 & 0.99 & 0.8 & 0.74 & 0 \\
0.0001                 & 0.95 & 1 & 1 & 1 & 1 & 1 & 1 & 0.8 & 0.87 & 0 \\
0.00005                & 0.96 & 1 & 1 & 1 & 1 & 1 & 1 & 0.8 & 0.91 & 0 \\
0.00001                & 0.98 & 1 & 1 & 1 & 1 & 1 & 1 & 0.8 & 0.96 & 0 \\
0.000005               & 0.98 & 1 & 1 & 1 & 1 & 1 & 1 & 0.8 & 0.96 & 0 \\
0.000001               & 0.98 & 1 & 1 & 1 & 1 & 1 & 1 & 0.8 & 0.98 & 0 \\ \hline
\end{tabular}
\end{table}

Next, for single-change-point scenarios involving, the detection performance is summarized in Table \ref{tab:success rate M=1}. The proposed GS-PF method demonstrates largely stable success rates despite variations in $\alpha$, consistently achieving accuracies over $90\%$ except for the piece-wise stationary case, which indicates that the GS-PF method maintains consistently low annotation and Hausdorff errors in most cases, accurately identifying both the number and location of change points. In contrast, while the GLASSO method can identify change points in constant and benchmark cases with relatively high success rates, it fails in symmetric and asymmetric cases. This is because the GLASSO method tends to overestimate the number of change points, and spurious change points may be still kept and hence less accurate.

\begin{table}[t]
\centering
\tiny
\caption{Success rates of change point detection with $M=1$ for the two methods under different datasets and $\alpha$ values.}
\label{tab:success rate M=1}
\begin{tabular}{ccccccccccc}
\hline
\multirow{2}{*}{$\alpha$} & \multicolumn{2}{c}{Constant} & \multicolumn{2}{c}{Symmetric} & \multicolumn{2}{c}{Asymmetric} & \multicolumn{2}{c}{Piece-wisely stationary} & \multicolumn{2}{c}{Benchmark} \\ \cline{2-11} 
& GS-PF & GLASSO & GS-PF & GLASSO & GS-PF & GLASSO & GS-PF & GLASSO & GS-PF & GLASSO \\ \hline
0.05                   & 0.97  & 0.98  & 1  & 0  & 1 & 0  & 0.48  & 0.52  & 1  & 1 \\
0.01                   & 0.96  & 0.98  & 1  & 0  & 1 & 0  & 0.48  & 0.52  & 1  & 1 \\
0.005                  & 0.96  & 0.98  & 1  & 0  & 1 & 0  & 0.48  & 0.52  & 1  & 1 \\
0.001                  & 0.95  & 0.98  & 1  & 0  & 1 & 0  & 0.48  & 0.52  & 1  & 1 \\
0.0005                 & 0.95  & 0.98  & 1  & 0  & 1 & 0  & 0.48  & 0.52  & 1  & 1 \\
0.0001                 & 0.91  & 0.98  & 1  & 0  & 1 & 0  & 0.47  & 0.52  & 1  & 1 \\
0.00005                & 0.91  & 0.98  & 1  & 0  & 1 & 0  & 0.47  & 0.52  & 1  & 1 \\
0.00001                & 0.91  & 0.98  & 1  & 0  & 1 & 0  & 0.47  & 0.52  & 1  & 1 \\
0.000005               & 0.91  & 0.98  & 1  & 0  & 1 & 0  & 0.47  & 0.52  & 1  & 1 \\
0.000001               & 0.90  & 0.98  & 1  & 0  & 1 & 0  & 0.46  & 0.52  & 1  & 1 \\ \hline
\end{tabular}
\end{table}

Finally for the most complicated scenarios with multiple change points in functional sequences, the performance are presented in Table \ref{tab:success rate M=5}. The GS-PF method consistently demonstrates a high success rate in detecting change points, particularly achieving $100\%$ accuracy for symmetric, asymmetric, and benchmark cases. Although a slight decrease in success rate is observed as $\alpha$ decreases in constant and piecewise-stationary cases, it is anticipated as the signal to noise may not be sufficiently strong to identify change points. A lower FDR level implies a more stringent criterion for rejecting the null hypothesis in multiple testing situations, leading to some $p$-values just above $\alpha$, not triggering a rejection and potentially underestimating the true number of change points. Enhancing the discrepancy in the mean function before and after the change point would result in much smaller $p$-values compared to $\alpha$, rendering the method less sensitive to $\alpha$ variations. Conversely, the GLASSO method fails to detect any change points, yielding a $0$ success rate across all cases, due to its inability to accurately detect the number of change points. In summary, these findings support the superiority of the GS-PF method in accurately identifying both the number and location of multiple change points.

\begin{table}[t]
\tiny
\centering
\caption{Success rates of change point detection with $M=5$ for the two methods under different datasets and alpha values.}
\label{tab:success rate M=5}
\begin{tabular}{ccccccccccc}
\hline
\multirow{2}{*}{$\alpha$} & \multicolumn{2}{c}{Constant} & \multicolumn{2}{c}{Symmetric} & \multicolumn{2}{c}{Asymmetric} & \multicolumn{2}{c}{Piece-wisely stationary} & \multicolumn{2}{c}{Benchmark} \\ \cline{2-11} 
& GS-PF & GLASSO & GS-PF & GLASSO & GS-PF & GLASSO & GS-PF & GLASSO & GS-PF & GLASSO \\ \hline
0.05                   & 0.89  & 0  & 1  & 0  & 1 & 0  & 0.61  & 0  & 1  & 0 \\
0.01                   & 0.87  & 0  & 1  & 0  & 1 & 0  & 0.61  & 0  & 1  & 0 \\
0.005                  & 0.87  & 0  & 1  & 0  & 1 & 0  & 0.61  & 0  & 1  & 0 \\
0.001                  & 0.85  & 0  & 1  & 0  & 1 & 0  & 0.60  & 0  & 1  & 0 \\
0.0005                 & 0.85  & 0  & 1  & 0  & 1 & 0  & 0.60  & 0  & 1  & 0 \\
0.0001                 & 0.81  & 0  & 1  & 0  & 1 & 0  & 0.60  & 0  & 1  & 0 \\
0.00005                & 0.81  & 0  & 1  & 0  & 1 & 0  & 0.59  & 0  & 1  & 0 \\
0.00001                & 0.80  & 0  & 1  & 0  & 1 & 0  & 0.59  & 0  & 1  & 0 \\
0.000005               & 0.78  & 0  & 1  & 0  & 1 & 0  & 0.59  & 0  & 1  & 0 \\
0.000001               & 0.77  & 0  & 1  & 0  & 1 & 0  & 0.59  & 0  & 1  & 0 \\ \hline
\end{tabular}
\end{table}

\section{Real data analysis}
In this section, the proposed GS-PF method is applied to two real datasets, i.e., the central England temperature (CET) records and 12-lead electrocardiogram data for arrhythmia, respectively, compared with the GLASSO method. Specifically, to estimate the functionals, the FPCA technique is adopted in both datasets, implemented via the 'fdapace' package in R. The value of $K$ in \eqref{eq:basis expansion} is determined by the smallest number of components that explain at least $99\%$ of a sample's variance. Parameters $\lambda$ and $\eta$ are optimized using a grid search strategy, aiming to minimize the BIC derived from the given data. Additionally, the false discovery rate $\alpha$ is fixed at 0.01.

\subsection{The CET data}
\begin{figure}[H]
\centering
\includegraphics[width=12cm]{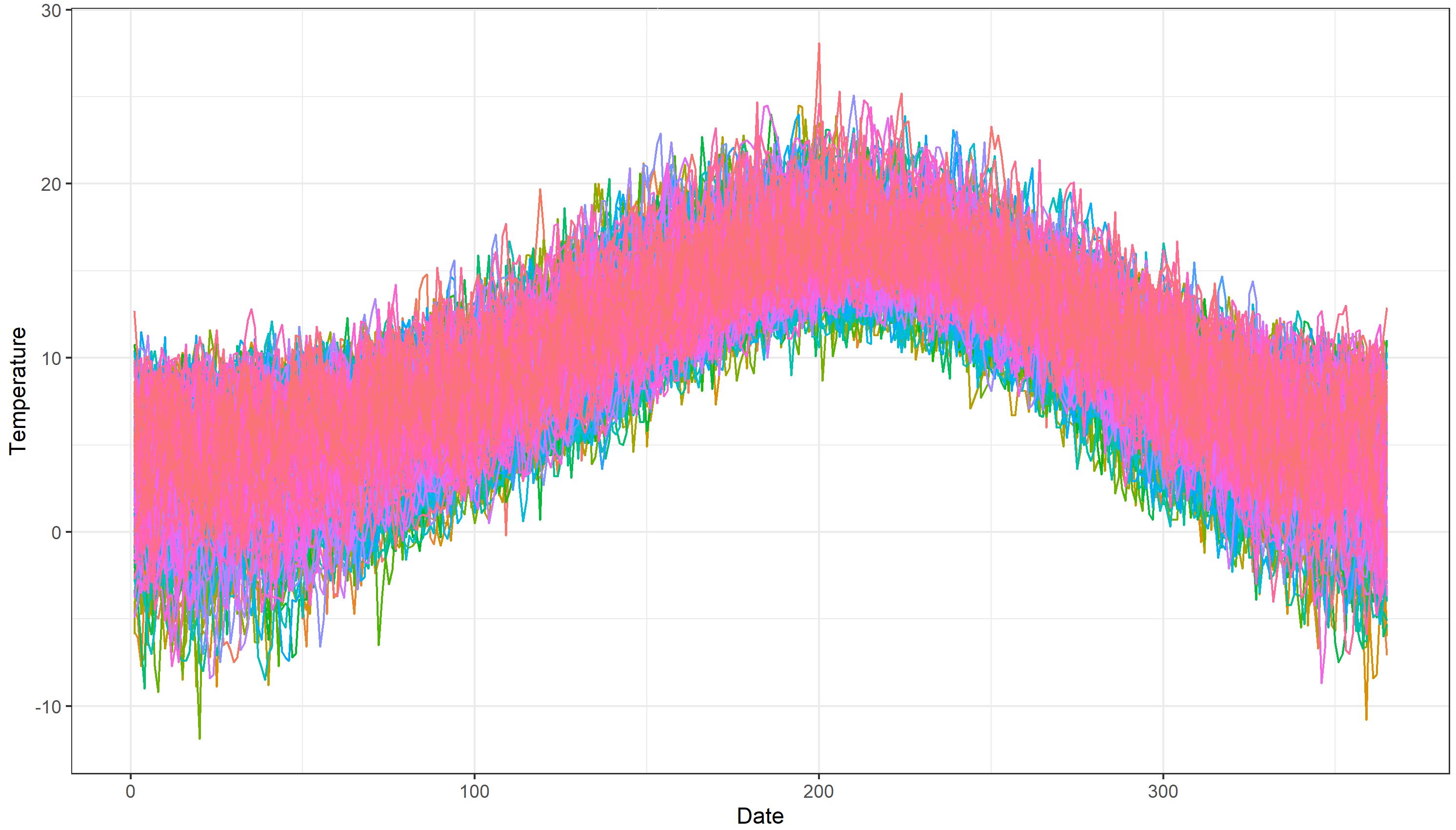}
\caption{Average daily temperatures in central England from 1772 to 2023.}
\label{fig:Line chart of CET}
\end{figure}

The CET data \cite{parker1992new} contains the average daily temperatures in central England from 1772 to 2023, which forms a functional data sequence of length $T=252$ with $d=365$ measurements in each function after deleting the data on $29^{th}$ Feb \footnote{https://www.metoffice.gov.uk/hadobs/hadcet/.}, illustrated in Figure \ref{fig:Line chart of CET}. The sequences fluctuate around a fixed trend, and accordingly the changes in their mean functions appear to be mere upward or downward shifts, while the overall trend remains constant.

\begin{figure}[H]
\centering
\includegraphics[width=12cm]{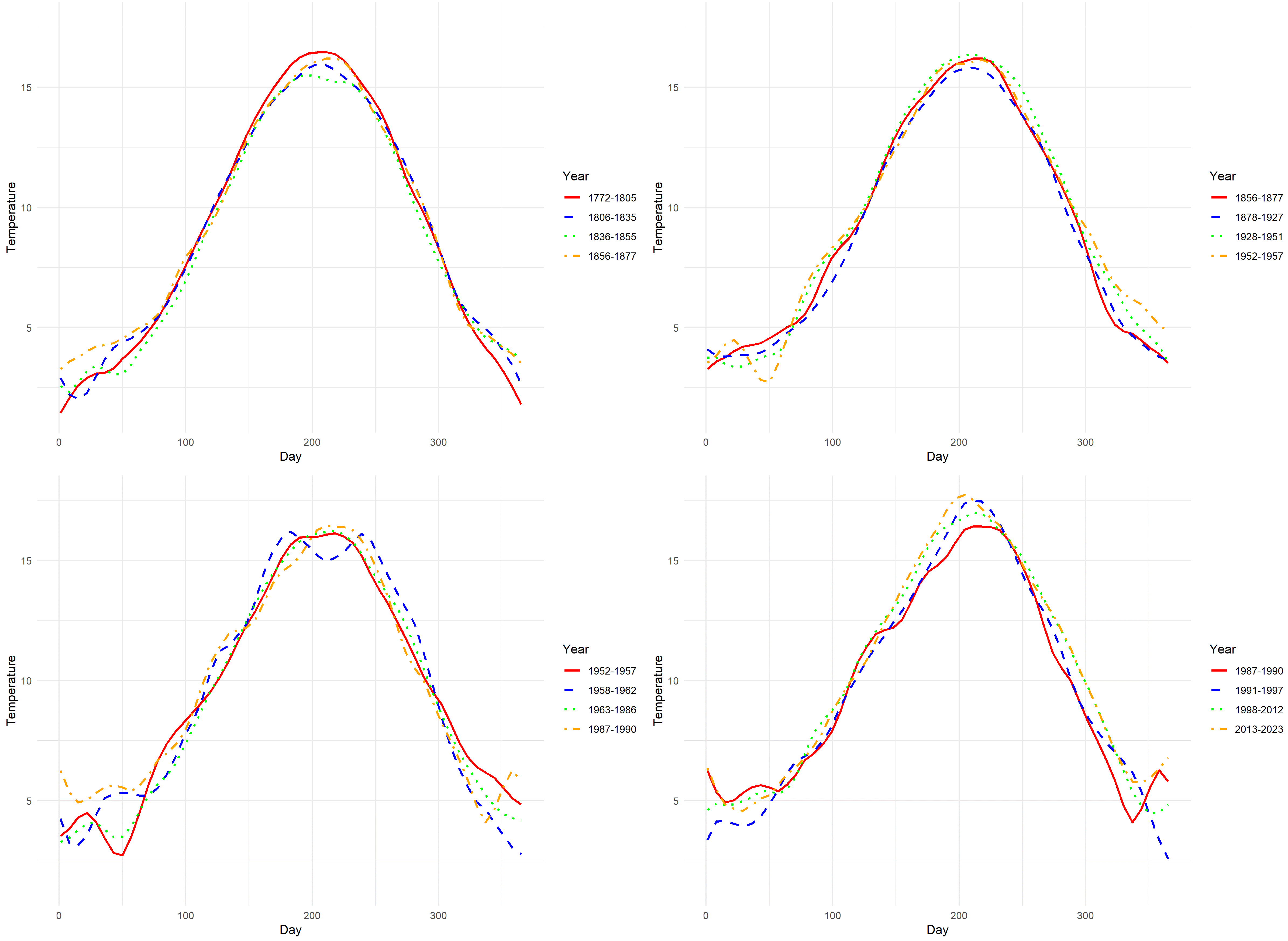}
\caption{Segment mean functions of CET data from 1772 to 2023.}
\label{fig:Line chart of CET mu}
\end{figure}

The proposed GS-PF method identifies 12 change points in the period from 1772 to 2023, as detailed in Table \ref{tab:cp of CET}.
These change points divide the dataset into 13 segments, each characterized by its own mean function. To better illustrate the changes at each change point, the 12 change points are grouped into four groups, each of which contains three change points, as is shown in Figure \ref{fig:Line chart of CET mu}. The analysis reveals that the first and second groups exhibit small changes, which mainly occur at the beginning and end of each year. The third and fourth groups, however, display more significant changes, particularly in the years 1952, 1958, and 1987. Notably, apart from a marked downward trend at the start of 1952, the mean function for 1958 has two peaks unlike the single peaks in other years, and there is a pronounced upward trend at the end of 1987. The change point detected in 1987 by the GS-PF method is consistent with findings in \cite{chen2023greedy} which employs greedy segmentation to identify multiple change points for functional data sequences. Additionally, the GS-PF method identifies other potential change points, such as the significant ones in 1952 and 1958, along with several points of smaller changes. In contrast, the GALSSO method identifies 19 change points. Notably, the change point detected in 1963 aligns with the GS-PF method. Additionally, the change points identified in 1837, 1876, 1986, and 2015 are in close proximity to those detected by the GS-PF method in 1836, 1878, 1987, and 2013, respectively. However, the change points detected by GLASSO tends to cluster, such as the three change points in 1820, 1821, and 1823, which may overestimate real change points. This pattern is also observed for change points in 1837 and 1838, as well as in 2021 and 2023.

\begin{table}[t]
\footnotesize
\centering
\caption{Change points of CET data detected by GS-PF and GLASSO method}
\label{tab:cp of CET}
\begin{tabular}{m{2cm} m{12cm}}
\hline
\multicolumn{1}{c}{Method} & \multicolumn{1}{c}{Detected Change Points} \\
\hline
\multicolumn{1}{c}{GS-PF} & 1806, 1836, 1856, 1878, 1928, 1952, 1958, 1963, 1987, 1991, 1998, 2013 \\
\hline
\multicolumn{1}{c}{GLASSO} & 1780, 1784, 1795, 1814, 1820, 1821, 1823, 1837, 1838, 1845, 1876, 1929, \\
& 1963, 1979, 1986, 2006, 2015, 2021, 2023 \\
\hline
\end{tabular}
\end{table}

\subsection{The 12-lead electrocardiogram data}
\begin{figure}
\centering
\includegraphics[width=12cm]{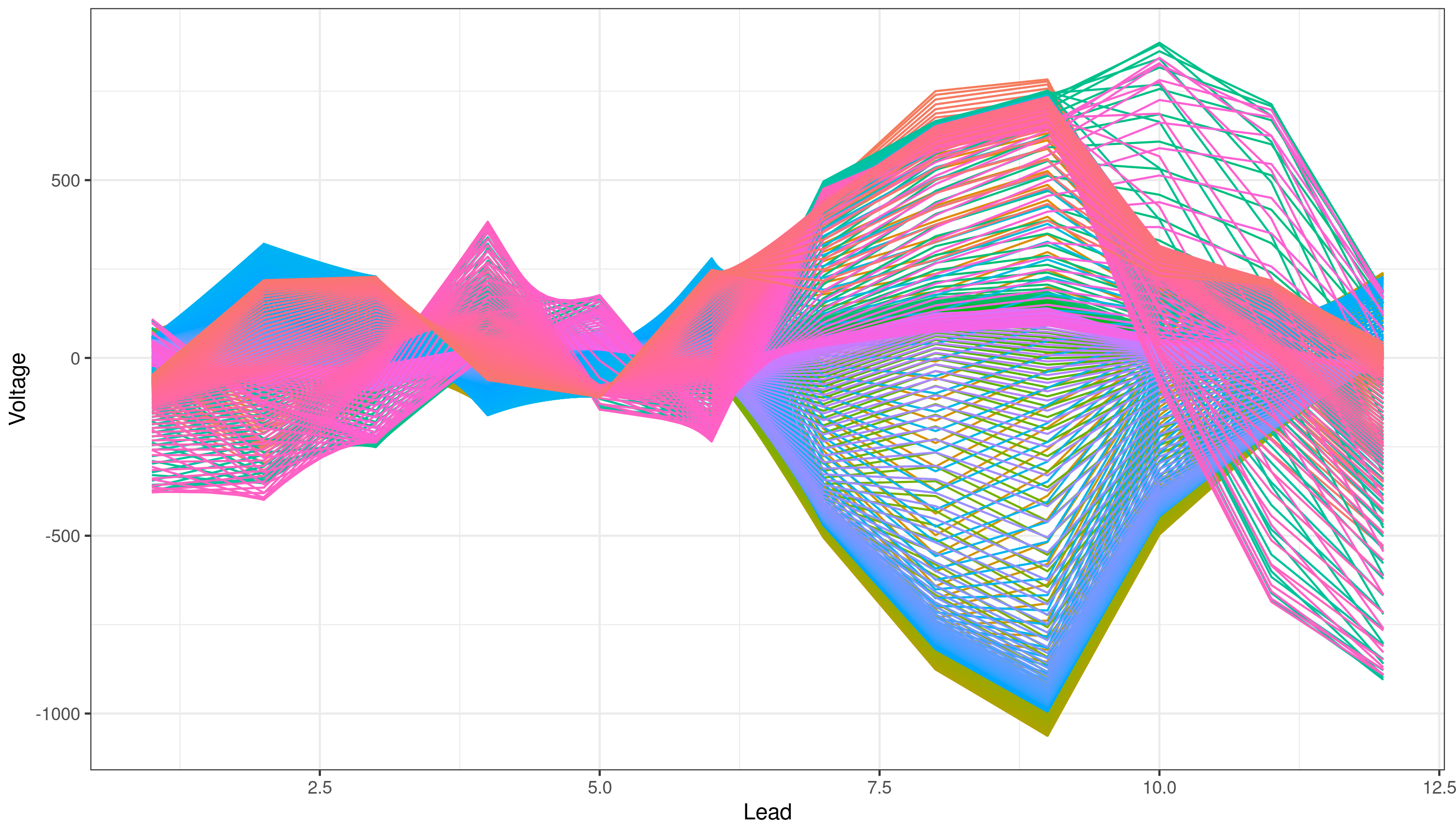}
\caption{Individual 12-lead electrocardiogram signals from 1 to 500.}
\label{fig:Line chart of ECG}
\end{figure}

The data for 12-lead electrocardiogram signals are collected for scientific studies on arrhythmia and other cardiovascular conditions \footnote{https://doi.org/10.6084/m9.figshare.11698521.}, which consist of 12-lead ECG data from 10646 patients, recorded at a high resolution of 500 Hz. The dataset offers a standardized 10-second sample for each individual by comprising 12-dimensional data points totaling 5000 samples. In our analysis, 500 consecutive samples are chosen from a random individual, which induces a functional data sequence of $T=500$ with $d=12$ measurements in each function. In contrast to the CET data, which exhibit fluctuations around a consistent trend, the ECG sequences display substantial variability, illustrated in Figure \ref{fig:Line chart of ECG}. This implies that the mean function of the ECG data exhibits significant differences between change points.

\begin{figure}[H]
\centering
\includegraphics[width=12cm]{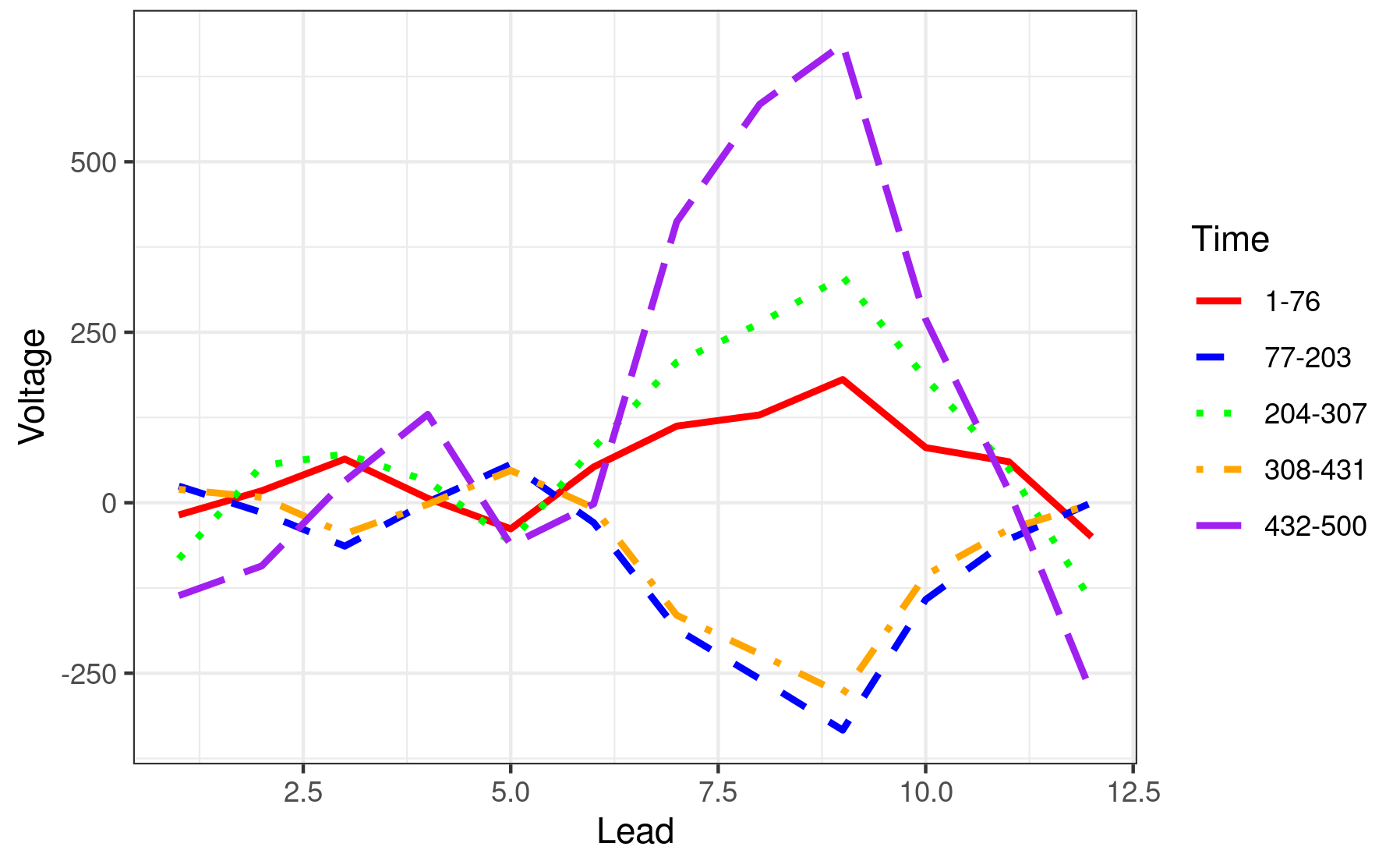}
\caption{Segment mean functions of ECG data from 1 to 500.}
\label{fig:Line chart of ECG mu}
\end{figure}

As is presented in Table \ref{tab:cp of ECG}, the proposed GS-PF method identifies 4 change points at time points, i.e., 77, 204, 308, and 432, respectively. As is anticipated, the differences between data segments before and after each change point are substantial, illustrated in Figure \ref{fig:Line chart of ECG mu}. The GLASSO method, however, detects a total of 228 change points, which are tightly clustered and grouped into seven groups. This high sensitivity to change points suggests that the GLASSO method may not achieve the change point accuracy, since it seems less reasonable that each sample in a series of continuous functions is identified as a change point, and it is implausible for experts to regard a series of continuous samples as change points in their field. This level of detection does not effectively assist experts and thus fails to serve the objective of efficient detection.

\begin{table}[t]
\footnotesize
\centering
\caption{Change points of ECG data detected by the GS-PF and GLASSO methods}
\label{tab:cp of ECG}
\begin{tabular}{ll}
\hline
Method & Detected Change Points \\
\hline
GS-PF & 77, 204, 308, 432 \\
\hline
GLASSO & 
\begin{tabular}{@{}l@{}}
Group 1: 23-40 (18 points) \\
Group 2: 64-130 (67 points) \\
Group 3: 202-213 (12 points) \\
Group 4: 241-271 (31 points) \\
Group 5: 297-354 (58 points) \\
Group 6: 429-439 (11 points) \\
Group 7: 469-499 (31 points) \\
\end{tabular} \\
\hline
\end{tabular}
\end{table}

\section{Conclusions}
In this article, a novel two stage GS-PF framework is proposed for detecting multiple change points in functional data sequences. The candidate change points are identified under the regression context using group penalty in the first stage, and a partial $F$-test procedure is then applied to remove redundant candidates in the second with a controlled false discovery rate. Asymptotic results are developed for the proposed method to guarantee a consistent detection. Further to improve the efficiency of the GS-PF method, a link parameter is used to generate candidate sets, which significantly reduces the number of candidate change points and speeds up the partial $F$-test. Numerical investigations on both simulated and real datasets validate the performance of the proposed GS-PF method across various scenarios. Our method is immune to three major challenges in the functional change point detection problem: accommodating a number of change points that may grow with sample size, operating without a prescribed minimum sample size between adjacent change points, and exhibiting robustness against asymmetric distributions and heavy-tailed sequences. The GS-PF framework may be further extended in various application scenarios for more complicated cases.

\bibliographystyle{chicago}
\bibliography{main}

\end{document}